\documentclass{epl}
\usepackage{tipa}

\begin{document}

\title{Plasmon assisted transmission of high dimensional orbital angular momentum entangled state}
\shorttitle{PAT of OAM entangled state}
\author{Xi-Feng Ren,Guo-Ping Guo\thanks{E-mail: \email:gpguo@ustc.edu.cn},Yun-Feng Huang,Chuan-Feng Li,Guang-Can Guo}
\shortauthor{Xi-Feng Ren \etal}

\institute{
  Key Laboratory of Quantum Information, University of Science and Technology of China, Hefei
230026, People's Republic of China }

\pacs{03.67.Mn}{Entanglement manipulation}

\pacs{42.50Dv}{Nonclassical states}

\pacs{71.36.+c}{Polaritons}

\maketitle

\begin{abstract}
We present an experimental evidence that high dimensional orbital
angular momentum entanglement of a pair of photons can be survived
after a photon-plasmon-photon conversion. The information of spatial
modes can be coherently transmitted by surface plasmons. This
experiment primarily studies the high dimensional entangled systems
based on surface plasmon with subwavelength structures. It maybe
useful in the investigation of spatial mode properties of surface
plasmon assisted transmission through subwavelength hole arrays.
\end{abstract}

Quantum entanglement is the foundation of quantum teleportation,
quantum computation, quantum cryptography, superdense coding, etc.
In recent years, the interest in high dimensional entangled states
is steadily growing. One advantage of using multilevel systems is
its promise to realize new types of quantum communication
protocols\cite{Bartlett00,Bourennane01}. Additionally, the usage of
multilevel systems provides a possibility to introduce very special
protocol, which cannot be implemented with qubits, such as quantum
bit commitment\cite{Langford04} and quantum coin
tossing\cite{Molina}. Another advantage is their possible
application in the fundamental tests of quantum mechanics. The most
popular approach to investigate higher dimensional systems relies on
the spatial modes of down-converted photons from spontaneous
parametric down conversion (SPDC). These down-converted photons can
be entangled in not only polarization, or spin angular momentum, but
also spatial modes, such as orbital angular momentum
(OAM)\cite{Mair01}, Hermite-Gaussian modes\cite{Ren042}. The spatial
modes entanglement occurs in an infinite-dimensional Hilbert
space\cite{Ren042,Arnaut00,Molina02,Torres04,
Mair01,Vaziri02,Vaziri03,Oem05}.

In metal films perforated with a periodic array of subwavelength
apertures, it has long been observed that there is an unusually high
optical transmission\cite{Ebbesen98}. Generally, it is believed that
surface plasmon (SP) in metal surface plays a crucial role in this
phenomenon, in which photons are first transformed into surface
plasmons and then back to photons\cite{Ebbesen98,Grupp00,Moreno}.
More recently, the experiments showing plasmon-assisted transmission
of entangled photons in polarization\cite{Alt} and in
energy-time\cite{energy} have attracted a lot of interest. These
experiments showed that the macroscopic surface plasmon
polarizations, a collective excitation wave involving typically
$10^{10}$ free electrons propagating at the surface of conducting
matter, can be treated as a single qubit. Apart from its fundamental
significance, this effect may provide another platform to transfer
entanglement between photons and condensed-matter materials.

So far, polarization properties of nanohole arrays have been studied
in many works\cite{Elli04,Gordon04,Altew05}. Then, how about the
spatial mode properties of nanohole arrays in the photon to plasmon
and back to photon process? The metal plates used in these
experiments always have special structure-usually like lattice.
Different from the conventional metal mirror, incident light is
converted into surface plasmon and the modes of surface plasmon are
correlated with these typical structures in metal plates. If the
light incidented on the metal plate has a transverse spatial
distribution itself, what will happen? Is there any modal dependence
on the geometry of the holes? What are the spatial modes of the
transmitted photons? In the previous experiments, it has been proved
that two dimensional entanglement such as polarization and
energy-time entanglement can be preserved in the surface plasmon
assisted transmission\cite{Alt,energy}. Our motivation is to see
whether the high dimensional spatial modes entanglement can be
survived in this photon to plasmon and back to photon process.

In this paper, we show that entanglement of OAM can be survived in
this plasmon assisted transmission process. The OAM entangled
qutrits are produced by type-I SPDC\cite{Mair01}. One photon of the
photon pair is incidented on the metal plate, undergoing a photon to
plasmon and back to photon process. We find the photon pairs after
the metal plate are still entangled in OAM. We may state that the
entanglement of OAM of photons can be transferred to surface
plasmons, and the information of spatial modes can be coherently
carried by surface plasmons.

\begin{figure}[b]
\centering
\includegraphics[width=6.0cm]{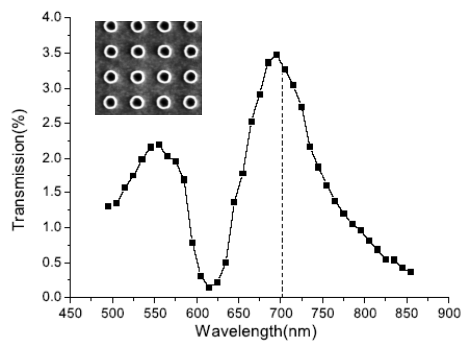}
\caption{Metal plate transmittance as a function of wavelength.
The dashed vertical line indicates the wavelength of 702nm used in
the experiment. Inset, scanning electron microscope picture of
part of our metal plate. After subsequently evaporating a 3-nm
titanium bonding layer and a 135-nm gold layer onto a 0.5-mm-thick
silica glass substrate, a Focused Ion Beam Etching system (FIB,
DB235 of FEB Co.) is used to produce cylindrical holes (200 nm
diameter) arranged as a square lattice (600 nm period). The total
area of the hole array is $30\mu m\times 30\mu m$ and it is
actually made up with four hole arrays of $15\mu m\times 15\mu m$
area for the technical reason.}
\end{figure}

\begin{figure}[b]
\centering
\includegraphics[width=8.0cm]{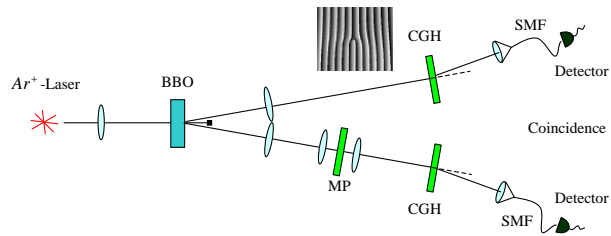}
\caption{(color online). Experimental set-up. After type-I
spontaneous parametric down conversion two lenses(focus 300mm) are
used to increase the collection efficiency. The metal plate(MP) is
placed between the twin lenses(focus 35mm) in the idler path. The
light incident on the metal plate has a diameter about $20\mu m$,
and thus cover several hundreds of holes. In signal and idler path,
computer generated hologram (CGH) and single mode fiber (SMF) are
combined to project the photons onto the different OAM states.
Inset, picture of part of a typical CGH ($+1$) with one fork in the
center. }
\end{figure}

Inset of Fig .1. is a scanning electron microscope picture of part
of our metal plate. The transmission spectrum is shown in Fig. 1.
The dashed vertical line indicates the wavelength of 702 nm used in
our experiment. The transmission of the metal plate at 702 nm is
about 3.2\%, which is much larger than the value of 0.55\% obtained
from classical theory\cite {ebbesen8}. This extraordinary optical
transmission is mediated by surface plasmon at the metal-substrate
interface\cite {Ebbesen98}. The metal plate is set between two
lenses of 35 \emph{mm} focal length, so that the light is normally
incident on the hole array with a cross sectional diameter of about
$20\mu m$ and covers several hundreds of holes.

\begin{figure}[b]
\centering
\includegraphics[height=5.0cm]{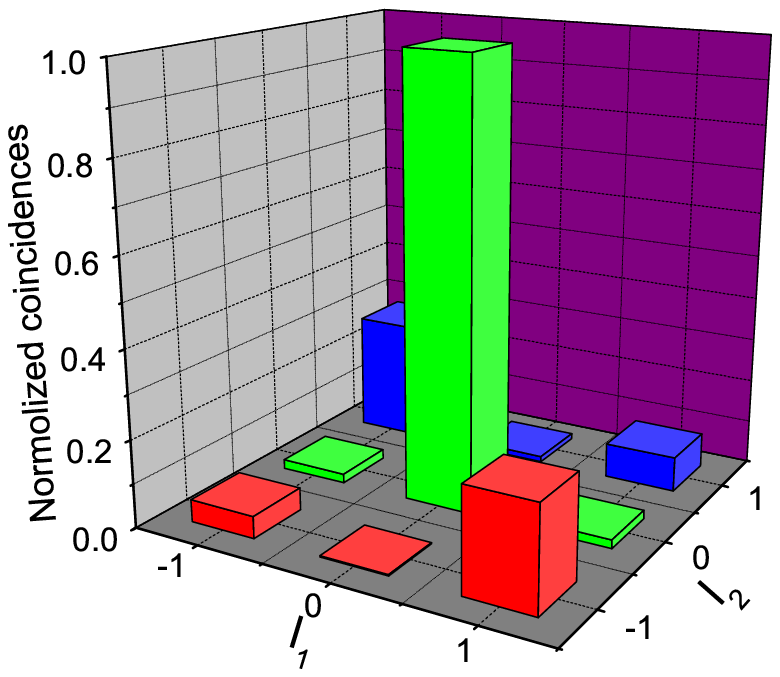}
\caption{(color online). Conversation of OAM when the metal plate
is moved out. Coincidence mode detections for signal photon and
idler photon in 9 possible combinations of orthogonal states are
performed. $l_1$ is the OAM of signal photon, and $l_2$ is the OAM
of idler photon. Within experimental accuracy, coincidences are
only observed in those cases where $l_1+l_2=0$.}
\includegraphics[height=5.0cm]{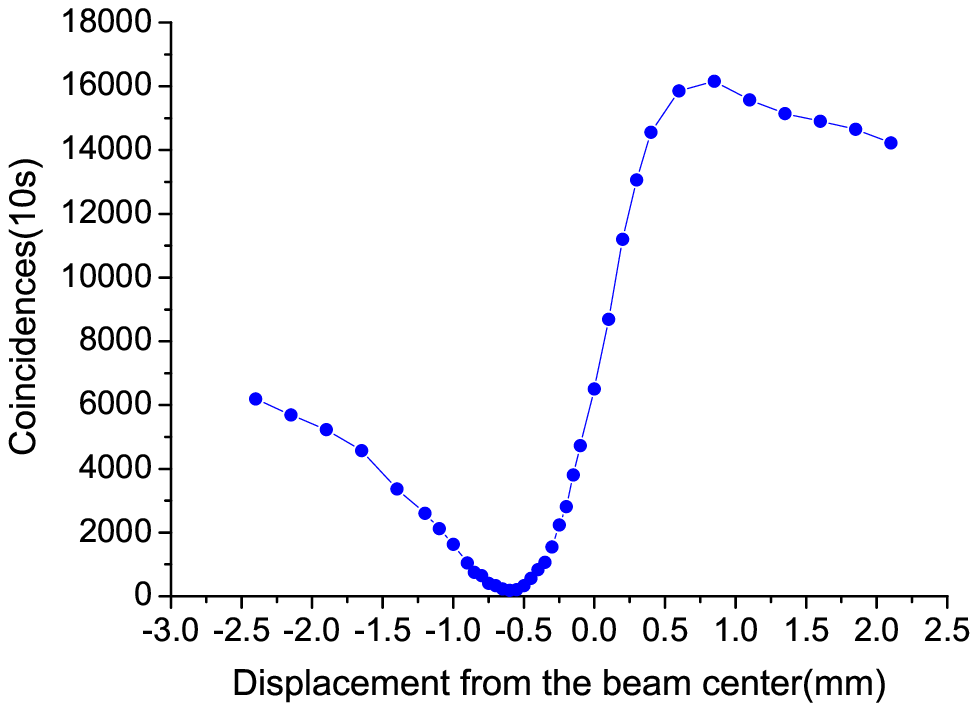}
\caption{Experimental evidence of entanglement of photon states with
phase singularities. The metal plate is moved out. Hologram in the
signal path is slightly displaced horizontally from the beam center.
Coincidences are recorded when we scan the hologram in the idler
path horizontally from one side to another side. Visibility is
$97.7\pm 0.1\%$.}
\end{figure}

\begin{figure}[b]
\centering
\includegraphics[height=5.0cm]{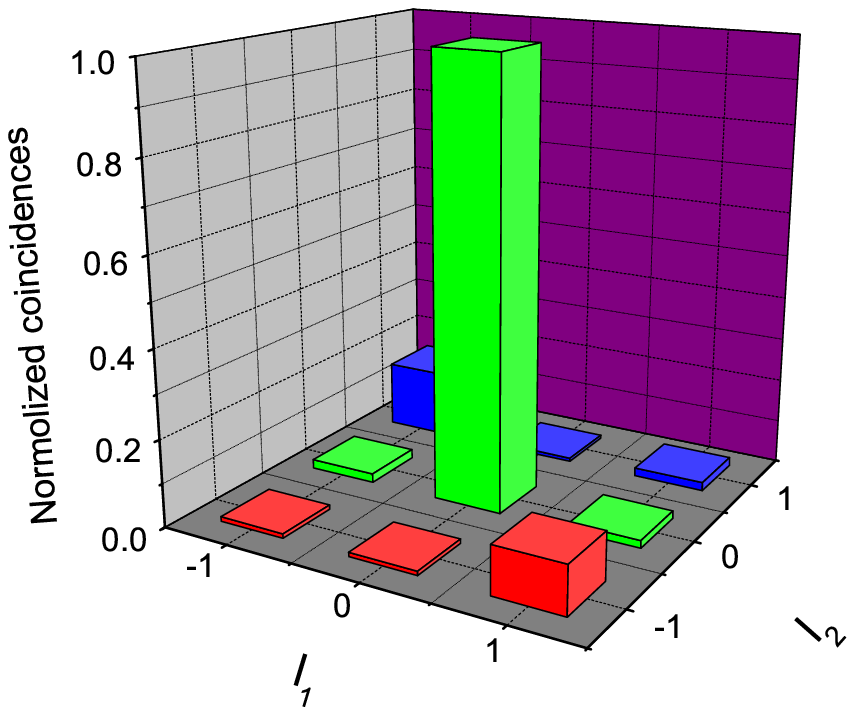}
\caption{(color online). Conversation of orbital angular momentum
when the metal plate is placed between the twin-lenses. Coincidence
mode detections for signal photon and idler photon in 9 possible
combinations of orthogonal states are also performed. As for the no
metal plate case, coincidences are only observed when $l_1+l_2=0$
within experimental accuracy.}
\includegraphics[height=5.0cm]{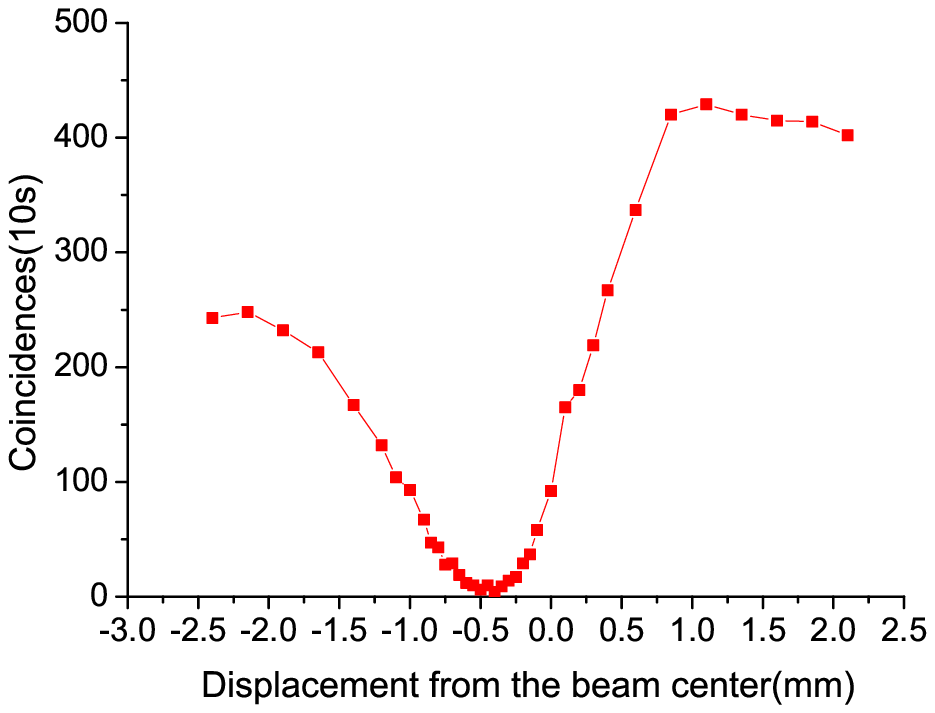}
\caption{Experimental evidence of entanglement of photon states with
phase singularities. The metal plate is placed between the
twin-lenses. Coincidences are recorded when we doing the same
operation as the case without metal plate. Visibility is $97.6\pm
0.5\%$.}
\end{figure}

In our experiment, high dimensional OAM entangled photon pairs are
produced via type-I spdc using a BBO crystal (beta barium borate)
of 3 mm thickness pumped by an argon ion laser operating at 351 nm
and having 400 mW light power. The pump light is in vertical
polarization. Both entangled photons have a wavelength of 702 nm
and  are in horizontal polarization.

The light fields of photons having OAM can be described by means of
Laguerre-Gaussian ($LG_p^l$) modes with two indices $p$ and
$l$\cite{Allen92}. The $p$ index identifies the number of radial
nodes observed in the transversal plane and the $l$ index describes
the number of the $2\pi$-phase shifts along a closed path around the
beam center. If the mode function is a pure LG mode with winding
number $l$ , then every photon of this beam carries an OAM of
$l\hbar $. This corresponds to an eigenstate of the OAM operator
with eigenvalue $l\hbar $\cite{Allen92}. If the mode function is not
a pure LG mode, each photon of this light is in a superposition
state, with the weights dictated by the contributions of the
comprised different $l$th angular harmonics. For the sake of
simplification, we can consider only LG modes with the index $p=0$,
and the $l$ index of the entangled photons varying from $-1$ to $1$.
Usually, we use computer generated holograms
(CGHs)\cite{ArltJMO,VaziriJOB} to change the winding number of LG
mode light. It is a kind of transmission holograms. Inset of Fig. 2.
shows part of a typical CGH($+1$) with a fork in the center.
Corresponding to the diffraction order $m$, the $l$ fork hologram
can change the winding number of the input beam by $\Delta l_m=ml$.
In our experiment, we use the first order diffraction light and the
efficiencies of  our CGHs are all about $30\%$. It is also important
to be able to both produce and analyze superposition states in a
chosen basis. A convenient method for creating superposition modes
is to use a displaced hologram\cite{VaziriJOB}, which is
particularly suitable for producing superpositions of $LG_0^l$ mode
with the Gaussian mode, which may also be seen as LG mode with $l=0$
.The Gaussian mode light can be identified using single-mode fibers
in connection with avalanche detectors. All other modes lights have
a larger spatial extension, and therefore cannot be coupled into the
single-mode fiber efficiently. High order LG modes ($l\neq 0$)
lights are identified using mode detectors consisting of CGH and
single-mode optical fiber. This mode detector can also be used to
identify the superposition mode by displacing the
CGH\cite{VaziriJOB,Langford04}.

The experimental setup is shown in Fig. 2. The Gaussian mode pump
light is focused on the BBO crystal where the entangled photon pairs
are produced. The photons are emitted from the crystal at an angle
of $6^\circ$ off the pump beam and are coupled into optical fiber
couplers via a lens ($f=300$mm) on each side. The metal plate is
placed between twin-lenses($f=35$mm) in the idler beam. The
twin-lenses are used to make the light incident on the metal plate
has a diameter about $20\mu m$ and thus cover several hundreds of
holes. CGH and single-mode fiber are combined to analyze the spatial
modes of the down converted photons. Silicon avalanche photodiode
(APD) single photon detectors (Perkin Elmer, SPCM-AQR-FC-16) are
used to record counts.

First, we move the metal plate out and measure the generated three
dimensional entangled state, or entangled qutrit state, as in the
previous works\cite{Mair01,Vaziri03}. The mode analysis is performed
in coincidence for all cases where mode detectors in signal and
idler path are prepared for analyzing LG modes -1, 0, 1
respectively. The measured coincidence pattern is shown in Fig. 3.
Within experimental accuracy, coincidences are only observed in
those cases where the sum of the OAM of the down-converted photons
is equal to the pump beam's OAM (here 0). This shows the
conservation of OAM. However, the absolute count rates of these
cases are not equal\cite{Mair01,Vaziri03,Langford04}.

To confirm entanglement, we have to demonstrate that the two photon
state is not just a mixture but a coherent superposition of product
states of the various gaussian and LG modes which obey OAM
conservation. There are several methods to prove entanglement, such
as testing the CHSH inequalities\cite{Vaziri02} and doing a quantum
tomography of the state\cite{Langford04}. Here we used the most
simple method which has been used in many
works\cite{Mair01,Vaziri03}. In signal path the CGH (here $+1$ or
$-1$) is slightly displaced horizontally form the beam center, while
in idler path we performed a scan of the mode of incoming photons by
moving the corresponding CGH horizontally from one side to another
side. An important distinction between coherent superposition and
incoherent mixture of gaussian and LG modes is that the latter
possess no phase singularity. If the state is in coherent
superposition, we will find a place of the CGH in the idler path
where the coincidences is zero. For example, if the generated state
is $(\left| 0,0\right\rangle +\left| -1,+1\right\rangle +\left|
+1,-1\right\rangle )/ \sqrt3$, and mode detector in signal path
detects the state $(\left| 0\right\rangle +\left| +1\right\rangle )/
\sqrt2$, then when the mode detector in idler path is adjusted to
detect the state $(\left| 0\right\rangle -\left| -1\right\rangle )/
\sqrt2$, we will get no coincidence. Define visibility as
$v=(C_{Max}-C_{min})/(C_{Max}+C_{min})$, where $C_{Max}$ is the
maximum of coincidence and $C_{min}$ is the minimum of coincidence.
If the state is a entangled state, $v$ will be $100\%$. The
experimental results shown in Fig.4. verify the correlation in
superposition bases LG ($l=\pm1$) and Gaussian modes. The visibility
is as high as $97.7\pm 0.1\%$ which can be viewed as a signature of
entanglement\cite{Mair01,Vaziri03}. This high visibility confirms
that the two photon state is not just a mixture but a entangled
state\cite{Mair01,Vaziri03}. Due to the non-linear relation between
the displacement of CGH and the superposition state detected by the
mode detector, the curve in Fig. 4 is  not symmetrical. Denoting a
state represented by the coincidence measurement of a photon at the
mode $m$ and a photon at the mode $n$ by $\left| m,n \right\rangle$,
we found that the generated entangled qutrit state can be written
as\cite{norm}:
\begin{eqnarray}
\left| \psi \right\rangle=(\left| 0,0\right\rangle +0.523\left|
-1,+1\right\rangle +0.486\left| +1,-1\right\rangle )/1.229.
\end{eqnarray}

Now, we put the metal plate between the twin-lenses in the idler
path and measure the output state of plasmon assisted transmitted
photons using the same method. The experimental results are shown in
Fig. 5. and Fig. 6. In order to demonstrate the entanglement, we
also measure the state in bases rotated in Hilbert space by
displacing the holograms. In the signal path, $+1$ (or $-1$)
hologram is displaced horizontally to detect superposition mode
$(a\left| 0\right\rangle +b\left| -1\right\rangle )/\sqrt{a^2+b^2}$
(or $(a\left| 0\right\rangle +b\left| +1\right\rangle
)/\sqrt{a^2+b^2}$), where $a$ and $b$ are real
numbers\cite{VaziriJOB}. While in the idler path the corresponding
hologram performs a scan of the mode of the incoming photons. If the
two photons are in entangled state, there will be a dip in the
coincidence pattern. The resulting coincidences are shown in Fig. 6.
The visibility is $97.6\pm 0.5\%$ which confirms the
entanglement\cite{Mair01,Vaziri03}. We can see there is a small
distance between this no-count place and the no-count place with no
metal plate (Fig. 4 and Fig. 6). The reason is that entangled state
is changed due to the different transmission efficiencies of
different modes light. These results show that plasmon assisted
transmitted photons are in the entangled qutrit state:
\begin{eqnarray}
\left| \psi' \right\rangle=(\left| 0,0\right\rangle +0.392\left|
-1,+1\right\rangle +0.332\left| +1,-1\right\rangle )/1.124.
\end{eqnarray}

So the OAM entanglement can be survived in the plasmon assisted
transmission. It is also noted that the degree of entanglement is
decreased after the photon to plasmon and back to photon process.
The reason is the unequal transmission efficiencies of different
modes lights. For photons carrying 0, -1, 1 OAM, the transmission
efficiencies are $3.25\pm 0.10 $\%, $1.51\pm 0.13 $\%, $1.82\pm 0.14
$\% respectively. We can directly get the state (2) from state (1)
by including these transmission efficiencies. If those transmission
efficiencies of different modes lights can be engineered properly,
these states may be concentrated to the maximally entangled qutrit
state with this kind of subwavelength metal optics\cite{Vaziri03}:
$(\left| 0,0\right\rangle +\left| -1,+1\right\rangle +\left|
+1,-1\right\rangle )/ \sqrt3$.

The phenomenon of unequal transmission efficiencies for different
OAM photons was also observed in a recent work\cite{ren06}. This
result might be explained by the theory of \cite{moreno,ruan}. In
these models, holes on the metal film are treated as waveguides, so
different modes lights can have different transmissivity as higher
modes lights decay faster in waveguide. However, these models cannot
be used to explain the conservation of orbital angular momentum
because the coherence among different modes may be destroyed in
these models. The relation between the spatial mode of photons and
the film perforation geometry is not yet given, which will be
investigated in our future work.

In conclusion, we proved experimentally that the OAM entanglement,
which is a high dimensional spatial modes entanglement, can be
coupled to surface plasmons. Our result may give us more hints to
the understanding of spatial mode properties of surface plasmon
assisted transmission.

\acknowledgments This work was funded by the National Fundamental
Research Program, National Nature Science Foundation of China
(10604052), the Innovation Funds from Chinese Academy of Sciences,
and the Program of the Education Department of Anhui Province (Grant
No.2006kj074A).


\begin{thebibliography}{0}
\bibitem{Bartlett00}  S. D. Bartlett, H. de Guise, and B. C. Sanders. quant-ph/0011080, (2000).

\bibitem{Bourennane01}  M. Bourennane, A. Karlsson, and G. Bj\"{o}rk. Phys.Rev.A. \textbf{64}, 012306, (2001).

\bibitem{Langford04}  N. K. Langford, R. B. Dalton, M. D. Harvey, J. L.
O'Brien, G. J. Pryde, A. Gilchrist, S. D. Bartlett, and A. G. White,
Phys.Rev.Lett. {\bf 93}, 053601, (2004).

\bibitem{Molina} G. Molina-Terriza, A. Vaziri, R. Ursin, and A. Zeilinger, Phys.Rev.Lett.
\textbf{94}, 040501 (2005).

\bibitem{Mair01}  A. Mair, A. Vaziri, G. Weihs, and A. Zeilinger, Nature
(London) {\bf 412}, 313, (2001).

\bibitem{Ren042}  X. F. Ren, G. P. Guo, J. Li, and G. C. Guo, Phys.Lett.A. {\bf 341}
81 (2004).

\bibitem{Arnaut00}  H. H. Arnaut and G. A. Barbosa, Phys.Rev.Lett. {\bf 85},
286, (2000).

\bibitem{Molina02}  G. Molina-Terriza, J. P. Torres, and L. Torner,
Phys.Rev.Lett. {\bf 88}, 013601, (2002).

\bibitem{Torres04}  J. P. Torres, A. Alexandrescu, and L. Torner,
Phys.Rev.A. {\bf 68}, 050301(R), (2003).

\bibitem{Vaziri02}  A. Vaziri, G. Weihs, and A. Zeilinger, Phys.Rev.Lett,
{\bf 89}, 240401, (2002).

\bibitem{Vaziri03}  A. Vaziri, J. W. Pan, T. Jenewein, G. Weihs, and A.
Zeilinger, Phys.Rev.Lett. {\bf 93}, 227902, (2003).

\bibitem{Oem05} S. S. R. Oemrawsingh, X. Ma, D. Voigt, A. Aiello, E. R. Eliel, G. W. 't Hooft, and J. P. Woerdman, Phys.Rev.Lett.
\textbf{95}, 240501 (2005).

\bibitem{Ebbesen98}  T.W. Ebbesen, H. J. Lezec, H. F. Ghaemi, T. Thio, and P. A. Wolff,  Nature 391, 667 (1998).

\bibitem{Grupp00}  D. E. Grupp, H. J. Lezec, T.W. Ebbesen, K. M. Pellerin, and T. Thio, Appl. Phys. Lett. 77, 1569 (2000).

\bibitem{Moreno}  L. Martin-Moreno, F. J. Garcia-Vidal, H. J. Lezec, K. M. Pellerin, T. Thio, J. B. Pendry, and T. W. Ebbesen, Phys. Rev. Lett. 86, 1114
(2001).

\bibitem{Alt}  E. Altewischer, M. P. van Exter and J. P. Woerdman Nature 418 304 (2002).

\bibitem{energy} S. Fasel, F. Robin, E. Moreno, D. Erni, N. Gisin and H. Zbinden, Phys. Rev. Lett. 94 110501 (2005).

\bibitem{Elli04} J. Elliott, I. I. Smolyaninov, N. I. Zheludev, and A. V. Zayats, Opt. Lett. 29, 1414 (2004).

\bibitem{Gordon04} R. Gordon, A. G. Brolo, A. McKinnon, A. Rajora, B. Leathem, and K. L. Kavanagh, Phys. Rev. Lett. 92, 037401
(2004).

\bibitem{Altew05} E. Altewischer, C. Genet, M. P. van Exter, and J. P. Woerdman, Opt. Lett. 30, 90 (2005).

\bibitem{ebbesen8}  H. A. Bethe, Phys. Rev. 66, 163 (1944).

\bibitem{Allen92}  L. Allen, M. W. Beijersbergen, R. J. C. Spreeuw, and J.
P. Woerdman, Phys.Rev.A. {\bf 45}, 8185, (1992).

\bibitem{ArltJMO} J. Arlt, K. Dholokia, L. Allen, and M. Padgett. J. Mod. Opt. 45
1231 (1998).

\bibitem{VaziriJOB} A. Vaziri, G. Weihs, and A. Zeilinger. J. Opt. B:
Quantum Semiclass. Opt 4 s47 (2002).

\bibitem{norm} The normalization of the states in this work is achieved by calculating the square root of the fraction given by the
coincidences for each LG mode divided by the total number of the
coincidences.

\bibitem{ren06} X. F. Ren, G. P. Guo, Y. F. Huang, Z. W. Wang, and G. C. Guo, Opt.
Lett. {\bf 31}, 2792, (2006).

\bibitem{moreno} L. Martin-Moreno, F. J. Garcia-Vidal, H. J. Lezec, K. M. Pellerin, T. Thio, J. B. Pendry, and T.W. Ebbesen, Phys. Rev. Lett. 86 1114 (2001).

\bibitem{ruan} Zhichao Ruan, and Min Qiu, Phys. Rev. Lett. 96 233901 (2006).
\end{thebibliography}
\end{document}